\documentclass[prb,twocolumn,superscriptaddress,showpacs,amsmath,amssymb]{revtex4-2}
\usepackage[colorlinks=true, linkcolor=blue, filecolor=blue, urlcolor=blue, citecolor=blue]{hyperref}
\usepackage{amsfonts}
\usepackage{bbm}
\usepackage{graphicx}
\usepackage{dcolumn}
\usepackage{bm}
\usepackage{soul}

\begin{document}

\title{Dissipation and dephasing in quantum Hall interferometers}

\author{Peng-Yi Liu}
\affiliation{International Center for Quantum Materials, School of Physics, Peking University, Beijing 100871, China}

\author{Qing-Feng Sun}
\email[]{sunqf@pku.edu.cn}
\affiliation{International Center for Quantum Materials, School of Physics, Peking University, Beijing 100871, China}
\affiliation{Hefei National Laboratory, Hefei 230088, China}

\begin{abstract}
In recent years, counter-intuitive results have shown that the quantum Hall edge states with topological protection can be dissipative.
In this paper, we point out that the non-equilibrium nature of edge states in quantum Hall interferometers leads to inevitable dissipation.
We consider a graphene interferometer operating in the integer quantum Hall regime and simulate the inelastic scattering that causes both dissipation and dephasing in the interferometer using non-equilibrium Green's function and virtual leads.
We describe the dissipation process with the numerical results of the spatial distribution of heat generation and the evolution of electron energy distribution.
In addition, with the enhancement of dephasing, a competition between Aharonov-Bohm interference and topologically protected quantized Hall plateaus is observed in the oscillations and fluctuations of the Hall resistances.
At a suitable dephasing strength, quantum Hall plateaus can be promoted by dephasing.
Our results not only give clues for the design of dissipation-free devices
but also provide a platform for studying the non-equilibrium relaxation and the dissipation mechanism of topological states.\\
\end{abstract}

\maketitle

\section{Introduction}
In the presence of a strong magnetic field, a two-dimensional electron
system exhibits zero longitudinal resistance and quantized Hall resistance, that is, the quantum Hall (QH) effect \cite{klitzing_new_1980}.
Due to the strong magnetic field, QH edge states form
close to the edge of the sample, which are characterized by chirality,
no backscattering, and robustness as topologically protected quantum states \cite{von_klitzing_40_2020}.
QH edge states have a long coherence length \cite{roulleau_direct_2008}, and the external magnetic field naturally provides electrons with controllable phase through the Aharonov-Bohm (AB) effect, which inspire people to design electronic interferometers in QH systems \cite{carrega_anyons_2021}.
Analogous to optical interferometers and with the use of
quantum point contacts (QPCs) as beamsplitters \cite{bocquillon_electron_2014}, QH interferometers such as Fabry-P\'{e}rot and Mach-Zehnder interferometers are designed theoretically and experimentally \cite{de_c_chamon_two_1997,halperin_theory_2011,robustness_Feldman_2022,time_Forghieri_2022,anyonic_Batra_2023,edge_Ihnatsenka_2023,van_wees_observation_1989,ji_electronic_2003,camino_aharonov-bohm_2005,zhang_distinct_2009,mcclure_edge-state_2009,ofek_role_2010,willett_magnetic-field-tuned_2013,single_Oksanen_2014,nakamura_aharonovbohm_2019}.
In recent years, QH interferometers have been studied in two-dimensional electron gas \cite{nakamura_aharonovbohm_2019,nakamura_direct_2020}, monolayer graphene \cite{wei_mach-zehnder_2017,ronen_aharonovbohm_2021,deprez_tunable_2021,zhao_graphene-based_2022}, bilayer graphene \cite{mahapatra_quantum_2022,fu_charge_2023,fu_aharonovbohm_2023} and other systems.
These high-performance QH interferometers have pioneered electron quantum optics \cite{bocquillon_electron_2014,bauerle_coherent_2018,jo_scaling_2022}
and have been used for the observation of fractional statistics \cite{willett_magnetic-field-tuned_2013,nakamura_direct_2020}.

Dephasing \cite{weak_Chakravarty_1986,phase_Stern_1990} in QH interferometers has attracted great research interest,
since reducing dephasing can improve the performance of QH interferometers
and the latter also provides a good platform to study dephasing \cite{roulleau_direct_2008,carrega_anyons_2021,jo_scaling_2022}.
Although identifying all the microscopic mechanisms of dephasing is difficult, many contributing factors have been found.
Within a single edge state, Coulomb interaction \cite{youn_nonequilibrium_2008} and collective spin excitations can cause dephasing \cite{roulleau_noise_2008}.
The coupling between co-propagating edge states \cite{roulleau_noise_2008,huynh_quantum_2012}, shot noise \cite{youn_nonequilibrium_2008,bhattacharyya_melting_2019} and edge reconstruction \cite{goldstein_suppression_2016,bhattacharyya_melting_2019} are also important factors of dephasing.
With the existence of these mechanisms, it is believed that dephasing
is widespread in QH edge states, despite the long coherence length.

Different from dephasing, since the discovery of the QH effect,
it has been widely believed that topologically protected QH edge states
are dissipation-free, and countless studies have mentioned this \cite{klitzing_the_1986,mohr_codata_2005,liu_quantum_2008,stern_non-abelian_2010,chang_experimantal_2013,xu_observation_2014,klevtsov_geometric_2015,zhao_tuning_2020,wang_quantum_2022,Chang_Colloquium_2023}.
Theoretically, however, dissipation due to the phonon emission is not forbidden in QH edge states \cite{slizovskiy_cooling_2017},
which can be enhanced by the inelastic forward scattering introduced
by defects and impurities \cite{halbertal_imaging_2017,zhang_dissipation_2020}.
In an experiment in 2019, thermal imaging of graphene using SQUID-on-tip was performed and heat generation along the QH edge states was observed, confirming the existence of dissipative transport in QH edge states \cite{marguerite_imaging_2019}.
This experiment is also supported by theoretical {researches} \cite{zhang_dissipation_2020,fang_thermal_2021,emergent_li_2024}.
Therefore, many problems related to QH edge states need to be reconsidered with the participant of dissipation urgently, such as the slight deviation between the Hall resistance and the theoretical value in experiments \cite{tzalenchuk_towards_2010}, the ``carrier cooling" problem \cite{seung_geo_thermoelectric_2013}, and the ``unknown channel for energy loss" \cite{le_sueur_energy_2010,krahenmann_auger-spectroscopy_2019}.
More importantly, the relationship between dissipation and topological protection of the QH edge states needs to be re-understood.

In this paper, we theoretically construct a graphene QH interferometer working in the integer QH regime, with the use of QPCs.
Based on the QH interferometer, we simulate the dissipation and dephasing of the QH interferometer by using non-equilibrium Greens's function and B\"{u}ttiker virtual leads \cite{buttiker_role_1986}.
On the one hand, we point out that as {long} as the QH edge states along the QH interferometer are in thermal non-equilibrium,
the dissipation and heat generation inevitably occur.
To eliminate dissipation, one needs to design a device that keeps the QH edge states in equilibrium all along while working.
We also calculate the spatial distribution of the heat generation
and the evolution of electrons energy distribution,
which can describe the dissipation process self-consistently.
On the other hand, by calculating the Hall resistance and its fluctuation, we observe a competition between AB oscillations and topologically protected quantized Hall resistance.
When the coherence length $L_{\phi}$ is longer than the {perimeter $L_{\rm{loop}}$} of the interference loop,
the visibility of the QH interferometer is high,
and the quantized Hall resistance is destroyed by AB oscillation.
With the increase of the dephasing,
the coherence length $L_{\phi}$ shortens.
When $L_{\phi}<{L_{\rm{loop}}}$, {AB oscillations are suppressed and quantized Hall plateaus appear}.
As a result, the dephasing promotes the appearance of the QH plateaus.
Under much stronger dephasing with the $L_{\phi}$ less than
the magnetic length $L_B$, QH plateaus are destroyed by the broadening plateau transition region, and the transport behavior tends to be classical \cite{pruisken_universal_1988}.
Our work will contribute to the design of QH interferometers, and we also provide a platform for studying the process of non-equilibrium relaxation and the dissipation mechanism of topological states.

The rest of the paper is organized as follows.
In Sec. \ref{II}, we present a theoretical model of a graphene QH interferometer and describe the methods and formulas we used.
In Sec. \ref{III}, we study the dissipation of the QH edge states in
the QH interferometer.
The effect of the dephasing on the interference pattern of resistance and the quantized Hall resistance {are} presented in Sec. \ref{IV}.
At last, Sec. \ref{V} concludes this paper.

\section{Theoretical Model} \label{II}

\begin{figure}
	\centering
	\includegraphics[width=\columnwidth]{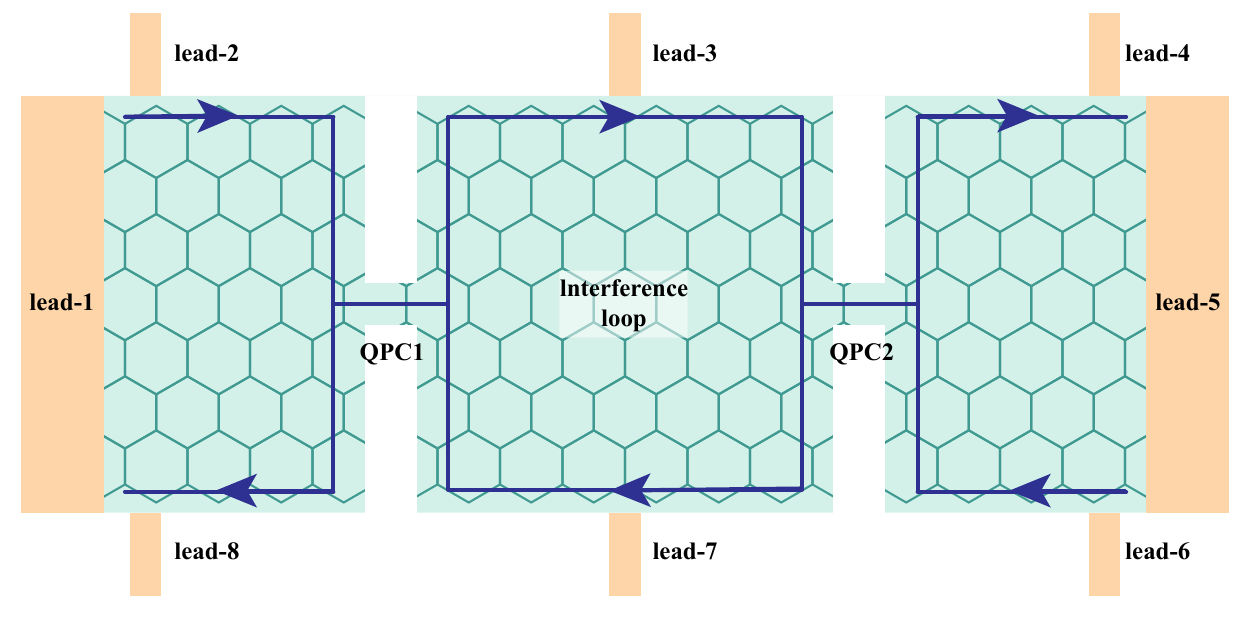}
	\caption{A schematic diagram of the graphene QH interferometer.
	The green-filled region represents a zigzag graphene nanoribbon and two QPCs are defined on it. Lead-1 and lead-5 act as the source and drain. Lead-(2-4) and lead-(6-8) are voltage probes. A QH edge state is shown by the dark blue arrow. The detailed dimensions of the QH interferometer are presented in Appendix \ref{A}.}
	\label{fig 1}
\end{figure}

To realize a QH interferometer, we require the establishment of two QPCs, which work as the beamsplitters \cite{carrega_anyons_2021}.
As shown in Fig. \ref{fig 1}, two QPCs are defined on the graphene, in which two narrow constrictions are introduced and the graphene becomes a Fabry-P\'{e}rot interferometer.
Lead-(1-8) are semi-infinite graphene leads contacted with the QH interferometer.
Lead-1 and lead-5 act as the source and drain,
and the rest six leads act as voltage probes.
With a strong perpendicular magnetic field,
the QH edge states form and the system becomes a QH interferometer.
For example, because of the narrow constrictions caused by QPCs,
one edge state is partially transmitted and
partially reflected at the QPCs.
We show the direction of an edge state in Fig. \ref{fig 1},
and a closed interference loop is formed between the two QPCs.

The QH interferometer, lead-(1-8){, and their coupling} can be described together by the graphene Hamiltonian,
\begin{equation}\label{eq1}
	H_G=\sum_i\epsilon_ic_i^{\dagger}c_i-\sum_{\langle ij \rangle}te^{\mathrm{i} \phi_{ij}}c_i^{\dagger}c_j,
\end{equation}
where $c_i$ ($c_i^{\dagger}$) is the annihilation (creation) operator of electrons at $i$ site,
$t=2.75{\rm{eV}}$ is the hopping energy,
$\epsilon_i=0.2t=0.55{\rm{eV}}$ is the on-site energy, and the magnetic field is introduced by the Peierls phase $\phi_{ij}=\int_i^j {\bm{A}}\cdot d{\bm{l}}/\phi_0$ \cite{long_disorder_2008}, with $\phi_0=\hbar/e$ the flux quantum.
We set the Fermi energy $E_F=0$.
Experimentally, the on-site energy $\epsilon_i$ can be adjusted by gate voltages, and the filling factor $\nu$ is
determined jointly by $E_F-\epsilon_i$ and the magnetic field.
When the Fermi energy lies in the gaps of the Landau levels, the physics of the edge states in the QH effect works.
For the rest of the paper, we set the unit of energy to ${\rm{eV}}$.

The dissipation sources (e.g., the electron-phonon interaction and impurities) inevitably exist in real systems and
can also convert the ordered energy of the electron motion 
into the disordered heat energy \cite{heat_Sun_2007,weak_Chakravarty_1986}.
We simulate the dissipation sources and dephasing by coupling B\"{u}ttiker virtual leads \cite{buttiker_role_1986,xing_influence_2008,fang_dephasing_2023,fang_thermal_2021,fang_thermald_2023} to
the QH interferometer.
{We couple virtual leads randomly to
the sites in the green and blue regions
in Fig. \ref{figa}(a) (see Appendix \ref{A}).
Each site in these regions has a probability $\eta=1/4$ to be coupled to a virtual lead.
Under a strong magnetic field, Landau levels are formed and the bulk is usually gapped. Therefore, there are no virtual leads in the gray regions in the bulk.}
As a result, electrons in the QH interferometer can go into and come back from the virtual leads with a loss of energy and phase memory.
Then, the total Hamiltonian of the system is given by \cite{fang_thermal_2021},
\begin{equation}\label{eq2}
{
H=H_G+\sum_{{i},k}\epsilon_{k}a_{ik}^{\dagger}a_{ik}
+\sum_{i,k}\left(t_{ik}a_{ik}^{\dagger}c_{i}+H.c.\right)},
\end{equation}
where the second and last terms represent the Hamiltonian of virtual leads
and their coupling to the QH interferometer, respectively.
{An electron in each virtual lead has a momentum $k$
and a continuous energy spectrum $\epsilon_k$.
$a_{ik}$ ($a_{ik}^{\dagger}$) is the annihilation (creation)
operator of electrons in the virtual lead-$i$ that coupled to site $i$
and $t_{ik}$ characterizes the coupling strength.
Here the coupling strength $t_{ik}$ is randomly either 0 or $t_k$
with the probability $1-\eta$ and $\eta$, respectively.
It is worth mentioning that here we assume that $t_{ik}$ takes only two values 0 or $t_k$. If $t_{ik}$ can take more values or a continuous distribution, the result remains unchanged, which is discussed in the Supplemental Material \cite{supp}.
For convenience, below we relabel all leads by $p\in \{1,2,3,...\}$.
$p\leq8$ represents real leads [lead-(1-8)], and $p>8$ for virtual leads.}

To simulate the Hall effect and heat generation of the system, we use the Landauer-B\"{u}ttiker formula to describe the relationship between the electric current, heat current, voltage, and temperature of both real and virtual leads \cite{datta_electronic_1995,fang_thermal_2021,xing_influence_2008},
\begin{equation}\label{eq3}
	\begin{split}
		&I_p=\frac{2e}{h}\sum_q \int T_{pq}(E)[f_p(E)-f_q(E)]{\rm d}E, \\
		&Q_p=-\frac{2}{h}\sum_q \int (E-\mu_p)T_{pq}(E)[f_p(E)-f_q(E)] {\rm d}E.
	\end{split}
\end{equation}
Here
$I_p$, $Q_p$, $\mathcal{T}_p$ and $f_p(E)=1/\{{\rm{exp}}[(E-\mu_p)/(k_B\mathcal{T}_p)]+1\}$ represent the electric current from the $p$-th lead to the QH interferometer, heat current from the QH interferometer to the $p$-th lead (i.e., the heat generation),
temperature of the $p$-th lead,
and Fermi distribution of the $p$-th lead, respectively.
We set the Fermi energy $E_F=0$ so that the chemical potential of the $p$-th lead is given by its voltage $\mu_p=eV_p$.
Besides, a spin degeneracy is assumed without considering the Zeeman splitting and spin excitation for simplicity.
$T_{pq}(E)$ is the transmission coefficient of electrons with energy $E$ from the lead-$p$ to lead-$q$.

To calculate the transmission coefficient, we use the non-equilibrium Green's function and $T_{pq}(E)={\rm{Tr}}\left({\bm{\Gamma}}_p {\mathbf{G}}^r {\bm{\Gamma}}_q {\mathbf{G}}^a\right)$ \cite{meir_landauer_1992}.
The retarded Green's function of the QH interferometer
is ${\mathbf{G}}^r(E)=[{\mathbf{G}}^a(E)]^{\dagger}=\left(E-\mathbf{H}_{\rm{QHI}}-\sum_p{\bm{\Sigma}}^r_p\right)^{-1}$, and the linewidth function ${\bm{\Gamma}}_p=\mathrm{i}[{\bm{\Sigma}}_p^r(E)-{\bm{\Sigma}}_p^a(E)]$.
$\mathbf{H}_{\rm{QHI}}$ is the Hamiltonian of electrons in the green region in Fig. \ref{fig 1}, {which is a part of $H_G$ in Eq. (\ref{eq1})}.
${\bm{\Sigma}}_p^r(E)=[{\bm{\Sigma}}_p^a(E)]^{\dagger}$ is the retarded self-energy due to the coupling of the $p$-th leads.
For virtual leads, the self-energy ${\bm{\Sigma}}_p^r=-\mathrm{i}\Gamma_d/2$ is energy-independent, with $\Gamma_d=2\pi\rho t_k^2$ and $\rho$ the density of states of virtual leads \cite{xing_influence_2008}.
$\Gamma_d$ represents the strength of dephasing and can be converted into the coherence length $L_{\phi}$ [see Appendix \ref{A} and Fig. \ref{figa} (b)].
For real leads, we use a numerical method to obtain its self-energy \cite{lee_simple_1981,sancho_highly_1985}.
In addition, with the Green's function, we can further calculate the energy distribution of electrons at $i$ site, which is given by the ratio of local electron density to the local density of states \cite{fang_thermal_2021},
\begin{equation}\label{eq4}
	F_i(E)=\frac{n_i(E)}{{\rm{LDOS}}_i(E)}.
\end{equation}
By definition, the electron density is related to the lesser Green's function $n_i(E)=-\frac{\mathrm{i}}{2\pi}{\mathbf{G}}_{ii}^<(E)$ and the local density of states is given by ${\rm{LDOS}}_i(E)=-\frac{1}{\pi}{\rm{Im}}{\mathbf{G}}_{ii}^r(E)$.
And the lesser Green's function can be calculated according to the Keldysh formulation \cite{fang_thermal_2021,haug_quantum_2008}, ${\mathbf{G}}_{ii}^<(E)=-\sum_p f_p(E){\mathbf{G}}_{ip}^r(E)[{\bm{\Sigma}}_p^r(E)-{\bm{\Sigma}}_p^a(E)]{\mathbf{G}}_{pi}^a(E)$.

With the discussion above, we can calculate the heat generation, resistances, and distribution of electrons once we have determined the boundary conditions of Eq. (\ref{eq3}).
For the rest of the article, we consider that our system has a good heat exchange with a huge environment, so we set all leads at the same temperature $\mathcal{T}_p=\mathcal{T}$ \cite{fang_thermal_2021}.
Then, Eq. (\ref{eq3}) can be reduced with small bias {voltages and low temperature} \cite{fang_thermal_2021},
\begin{eqnarray}
		I_p &= & \frac{2e^2}{h}\sum_{q\neq p} T_{pq}(0)(V_p-V_q), \label{eq5} \\
		Q_p &= & -\frac{2e^2}{h}\sum_{q\neq p} T_{pq}(0)\left(V_pV_q-\frac{1}{2}V_p^2-\frac{1}{2}V_q^2\right). \label{eq6}
\end{eqnarray}
We consider a small bias between lead-1 and lead-5, i.e., $V_1=V_0$ and $V_5=0$.
The approximation from Eq. (\ref{eq3}) to Eqs. (\ref{eq5}) and (\ref{eq6})
works well with
$(k_B \mathcal{T})^2\ll \Delta E\times eV_0\ll (\Delta E)^2$ \cite{fang_thermal_2021},
where $\Delta E$ is the gap between the Landau levels that we care about.
We guarantee this condition in all calculations and it is also accurate in typical experiments \cite{anomalous_Biswas_2024}.
The currents of other leads are set to be zero $I_{p\neq 1,5}=0$, {since electrons will not leave the system in a real dissipation process and lead-(2-4, 6-8) act as voltage probes}.
With these conditions, the voltage $V_{p}$ in the lead-$p$ can be solved from Eq. (\ref{eq5}) directly because Eq. (\ref{eq5}) is a linear equation.
Then we can obtain $I=I_{1}=-I_{5}$ and $Q_{p}$ from Eqs. (\ref{eq5}) and (\ref{eq6}) straightforwardly.
At last, we can obtain resistances $R_{pq}=(V_p-V_q)/I$, and the heat generation of dissipation is given by $Q_{p>8}$.


\section{Dissipation of the QH edge states}\label{III}

In this section, we discuss the dissipation of the QH edge state
along the edge of the QH interferometer.
For simplicity, we fix the magnetic field $B=0.03$
(in the unit of $4\hbar/3\sqrt{3}ea^2$)
to make sure there is only one edge state,
where $a$ is the length of the carbon-carbon bond in graphene.
The width of the QH edge state can be estimated by magnetic length $L_B=\sqrt{\hbar/eB}\approx6.58a$.
It is close to the width of the constriction $M=5a$ introduced by QPCs (see Appendix \ref{A}), so two QPCs can work as the beamsplitters.
The specific parameters we used in the calculation are given in Appendix \ref{A}.

\subsection{Heat generation}

\begin{figure*}[!htb]
	\centerline{\includegraphics[width=2\columnwidth]{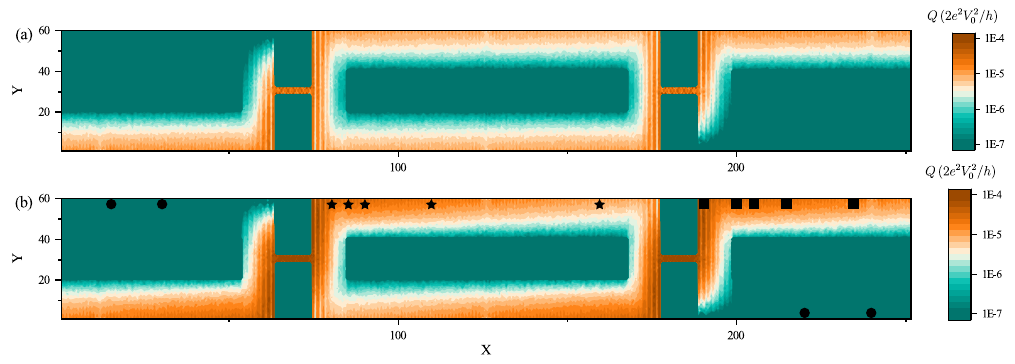}}
	\caption{The averaged dissipative heat generation of different sites.
		Each site is described by a coordinate $(x,y)$ with the origin located
in the bottom left corner, see Appendix \ref{A}.
		We choose a small dissipation strength $\Gamma_d=0.02$ in (a) and a large strength $\Gamma_d=0.1$ in (b).
		$B=0.03$ and other parameters are given in Appendix \ref{A}.
	The power of heat generation, i.e., the magnitude of the heat current of the virtual leads, is indicated in different colors, see the color bar.
The black marks in (b) give the positions that
we show the energy distribution of the electrons in Fig. \ref{fig 3}.
	}
	\label{fig 2}
\end{figure*}

With Eqs. (\ref{eq5}) and (\ref{eq6}) and the boundary conditions,
we can calculate the heat current from the QH interferometer
to each virtual lead, which is the heat generation due to the dissipation.
The power of {the} heat generation is the magnitude of {the} heat current of virtual leads.
Here we take an average of 1000 random configurations
of virtual leads so that each site has an average heat generation.
The numerical results with two different strengths $\Gamma_d$,
are shown in Fig. \ref{fig 2}.

It can be seen that there is significant heat generation along the QH edge states, including in the whole interference loop, in the top right corner,
and in the bottom left corner.
This means that the dissipation occurs
when electrons propagate along the chiral QH edge state.
This is contrary to the general belief
that topologically protected edge states are dissipationless.
{But} this is consistent with the experiment that the temperature increase caused by dissipation is observed downstream along the QH edge state \cite{marguerite_imaging_2019}.
Although we set $V_0$ is small, our results show that $Q_p/V_0^2$ will remain constant{s} under a fixed $\Gamma_d$, which implies
the occurrence of the dissipation even if for a very small bias $V_0$.

For a small $\Gamma_d=0.02$, as shown in Fig. \ref{fig 2}(a),
the heat generation along the QH interferometer loop almost does
not decay with the increase of the propagation distance,
which is consistent with the imaging experiment \cite{marguerite_imaging_2019}.
For a large $\Gamma_d=0.1$, as shown in Fig. \ref{fig 2}(b),
the heat generation near the QPCs
and along the QH edge are both larger than Fig. \ref{fig 2}(a).
In the parameters of Fig. \ref{fig 2}(b),
the Hall resistance $R_{37}=1.0098(h/2e^2)$ which is well close to
the resistance quantize,
which means that topological protection plays its role
and the chiral QH edge state still exists in our system
with the introduction of virtual leads.
Therefore, the results in Fig. \ref{fig 2} demonstrate a
counter-intuitive fact that topological protection cannot
prevent the dissipation of quantum states.
Even if topological protection exists and backscattering is prohibited, dissipative heat production of the QH edge state can still occur.

Next, we focus on the conditions under which dissipation does not occur.
It is not a surprise that {the} QH bulk has no dissipative heat generation because it is insulated and does not participate in transport on the QH plateau regime.
The QH edge state in the top left and bottom right corners of
the QH interferometer does not dissipate as well.
In fact, in the top left (bottom right) corner,
electrons on the QH edge state are directly from the lead-1 (lead-5)
and are not affected by QPCs,
so their energy distribution remains an equilibrium
with the chemical potential being the same as that of the lead-1 (lead-5)
as shown in Fig. \ref{fig 3}(a).
This ensures that no matter how large the $\Gamma_d$ is,
there is also no dissipative heat generation
in the top left and bottom right corners of our system.
That is, the dissipation does not occur
only if the electrons are in equilibrium.

In addition, the energy conservation in Fig. \ref{fig 2} can be verified.
The total power input to the system is the current flowing from the source (lead-1) to the drain (lead-5) multiplied by the bias $W=I_1V_0$.
And the total power going out of the system is the sum of dissipative heat generation of the whole QH interferometer and all contacts, that is, the heat current goes out of all leads $Q_{\rm{total}}=\sum_p Q_p$ (both real and virtual leads).
If the dissipation within the QH interferometer is reduced,
the dissipation at contacts will increase.
$W=Q_{\rm{total}}$ holds true for any $\Gamma_d$.
The QH interferometer also outputs entropy to the environment (virtual leads).
Since we theoretically set the temperature of the system to be the same,
the entropy generated by the edge state dissipation per unit time is $\sum_{p>8}Q_p/\mathcal{T}$ \cite{marguerite_imaging_2019},
which comes from the relaxation process from the non-equilibrium distribution to the equilibrium one, as we discussed in the next subsection.

\subsection{Energy distribution along the QH edge states}

\begin{figure}[!htb]
	\centerline{\includegraphics[width=\columnwidth]{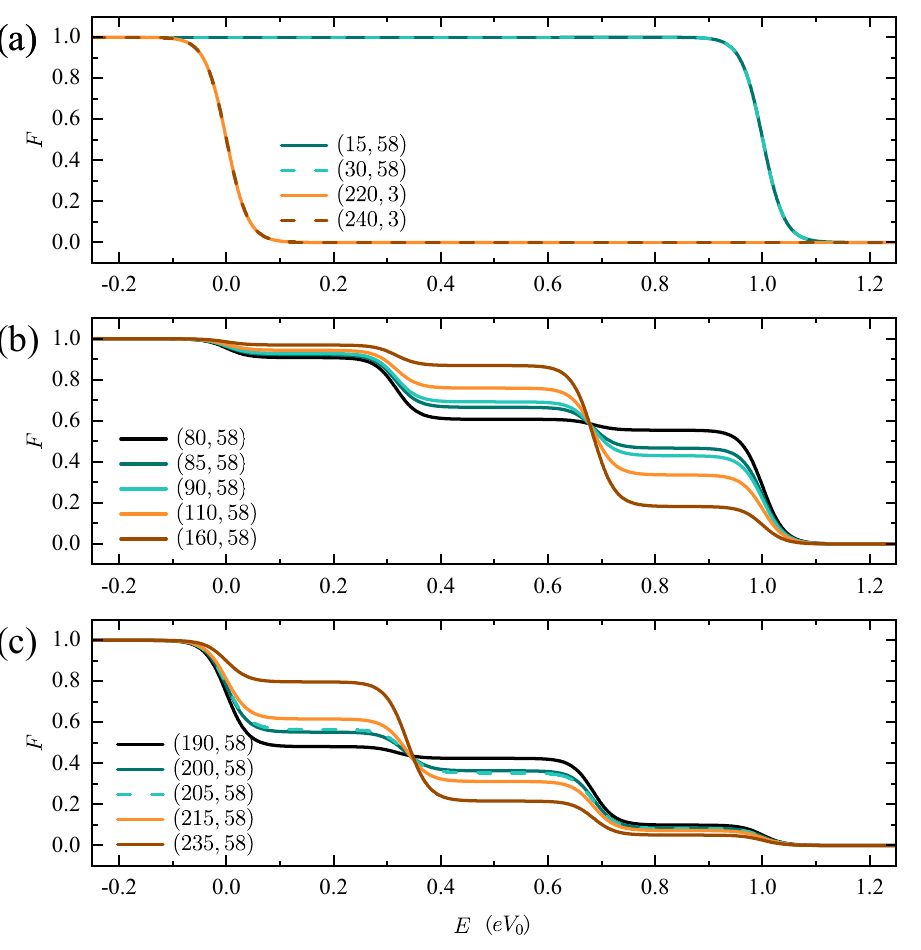}}
	\caption{The energy distribution $F$ of the electrons at different positions as marked in Fig. \ref{fig 2}(b).
		(a) The energy distribution of the electrons in the top left and the bottom right corners of the QH interferometer, shown as black dots in Fig. \ref{fig 2}(b).
		(b) The energy distribution of electrons in the top edge of the interference loop, represented by black stars in Fig. \ref{fig 2}(b).
		(c) The energy distribution of electrons in the top right corner of the QH interferometer, and their positions are represented by black squares in Fig. \ref{fig 2}(b).
		We set a temperature $k_b\mathcal{T}=0.02eV_0$, a strong dissipation $\Gamma_d=0.1$, and $B=0.03$.
		Other parameters are given in Appendix \ref{A}.
	}
	\label{fig 3}
\end{figure}

The energy distribution of electrons in QH edge states can be measured
by the current tunneling through a quantum dot, which is a kind of spectroscopy \cite{altimiras_non-equilibrium_2010,le_sueur_energy_2010}.
As we discussed in Sec. \ref{II}, we can use Eq. (\ref{eq4}) to calculate the energy distribution of electrons.
In Fig. \ref{fig 3}, we show the numerical results of the energy distribution at different positions as marked in Fig. \ref{fig 2}(b).
The equilibrium energy distribution [Fig. \ref{fig 3}(a)]
and the non-equilibrium distribution [Fig. \ref{fig 3}(b,c)]
that gradually evolves with the propagation of the edge state
are clearly shown.
When we calculate the energy distribution, we remove the virtual leads
near two QPCs (see Appendix \ref{A}), which made the calculation result more elegant on the one hand.
On the other hand, the dissipation sources that enhance the phonon emission (such as the single atom defect) in the experiment are usually distributed in the physical edge of graphene \cite{halbertal_imaging_2017,marguerite_imaging_2019,ronen_aharonovbohm_2021}, and there was less near the gate-defined constrictions.

As shown in Fig. \ref{fig 3}(a), in the top left and
bottom right corners of the QH interferometer,
electrons in {the} QH edge state are in equilibrium Fermi distributions,
since electrons come directly from the lead-1 and lead-5
without the backscattering.
The distribution $F$ of electrons in the bottom right corner
is the same as that in lead-5, where $F(E)=f(E)=1/[\exp(E/k_B\mathcal{T})+1]$.
The distribution $F$ in the top left corner
has the same chemical potential as lead-1,
so the distribution function is $F(E)=f(E-eV_0)$.
In equilibrium, low-energy states are fully filled,
and due to the Pauli exclusion principle, high-energy electrons cannot lose energy, even if dissipation sources are present.
This is consistent with the absence of dissipative heat generation
near the black dot in Fig. \ref{fig 2}(b),
and is also consistent with the prevailing view \cite{buttiker_absence_1988}.

When the equilibrium electrons propagating along the QH edge state
in the top left corner entering from lead-1 encounter
the left QPC, transmission and reflection occur with non-integer probabilities.
This results in a highly non-equilibrium double-step distribution of the transmitted electrons with $F(E) \approx f(E)+T[f(E-eV_0)-f(E)]$ \cite{buttiker_absence_1988,datta_electronic_1995,altimiras_non-equilibrium_2010,fang_thermal_2021}, where $T=0.5378$ is the transmission coefficient of a single QPC we applied.
Of course, the reflected electrons are also
in a non-equilibrium double-step distribution.
As the non-equilibrium electrons propagate forward
along {the} QH edge state in the interferometer,
high-energy electrons can lose energy through all kinds of dissipation mechanisms simulated by virtual leads.
Because the low-energy state is not fully filled,
nothing can stop the relaxation process
which causes the dissipative heat generation and entropy increase.
Even with the topological protection, as electrons along the edge state propagate forward, the high-energy electrons gradually decrease and the {low-energy} electrons gradually increase.
As shown in Fig. \ref{fig 3}(b) and (c), the energy distribution of electrons gradually evolves from non-equilibrium to equilibrium as some triple-steps.
In this process, the energy lost by the electrons flows into the environment in the form of a dissipative heat generation \cite{zhang_dissipation_2020,marguerite_imaging_2019}, which is what we observe in Fig. \ref{fig 2}.
It can be seen that less heat is generated at $(160,58)$ of Fig. \ref{fig 2}(b) than at the four black stars on the left because the high-energy electrons decrease with the energy relaxation, as shown in Fig. \ref{fig 3}(b).

Many previous experiments may have implied the dissipation caused
by such non-equilibrium relaxation.
One of them observed that the electron-electron interaction between different QH edge states can cause the relaxation of
non-equilibrium edge states.
But after the relaxation, an electron temperature lower than
theoretically expected was observed,
indicating an unknown channel for energy loss \cite{le_sueur_energy_2010,krahenmann_auger-spectroscopy_2019}.
Another experiment observed that the thermoelectric signal
of hot electrons is observed to decay with distance from a constriction,
indicating part of the heat is transferred out of the edge channel \cite{seung_geo_thermoelectric_2013}.
When interpreting these experiments, it was generally believed
that the dissipative mechanisms such as phonon emission were weak enough to be ignored.
However, both the thermal imaging experiment \cite{marguerite_imaging_2019} and our numerical results suggest
that dissipation is also an important energy destination
for non-equilibrium relaxation.

\begin{figure*}[!htb]
	\centerline{\includegraphics[width=2\columnwidth]{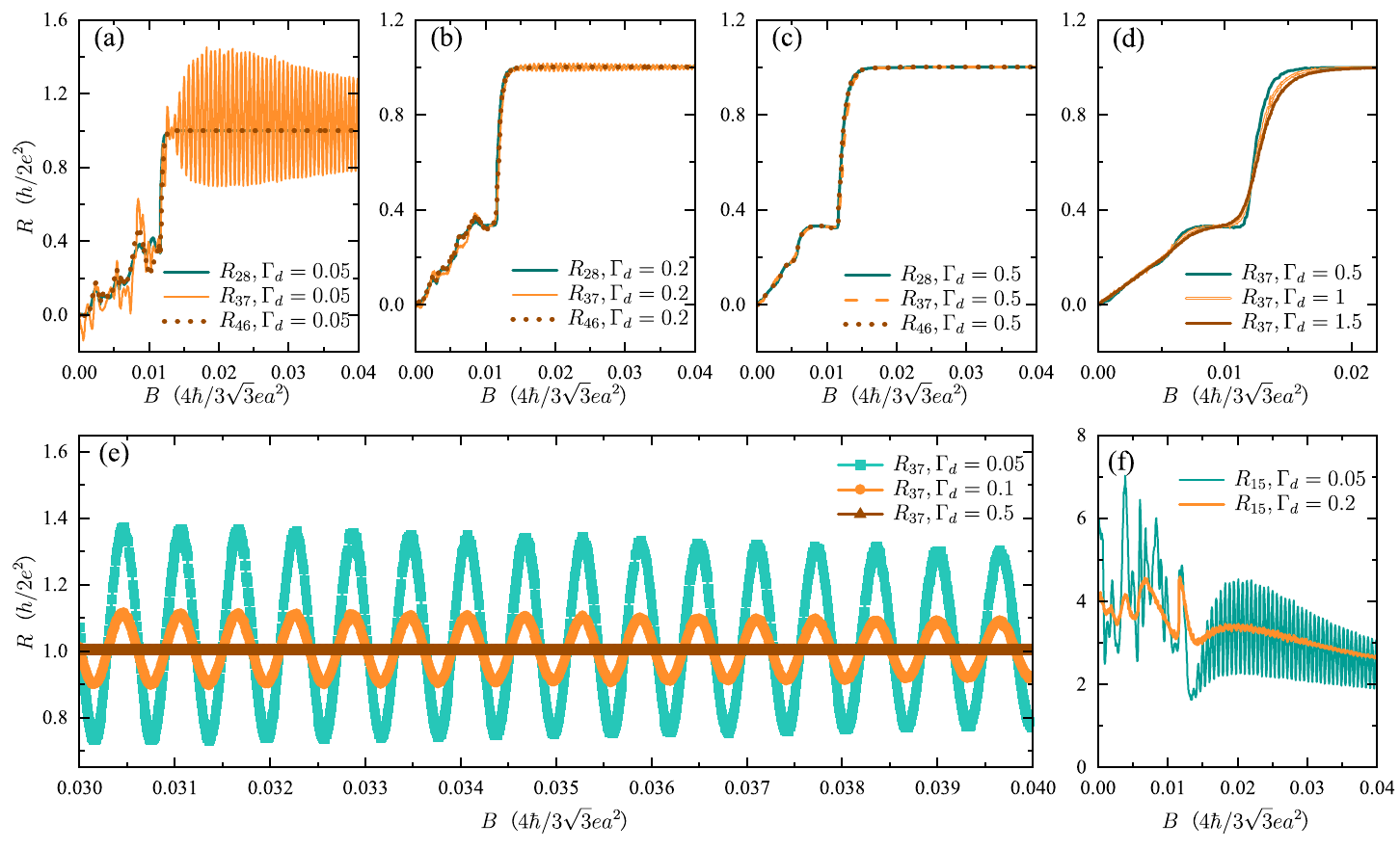}}
	\caption{(a-c) Hall resistances $R_{28}$, $R_{37}$, and $R_{46}$ versus the magnetic field $B$ at three different dephasing strength $\Gamma_d=0.05$, $0.2$, and $0.5{\rm{eV}}$.
	(d) $R_{37}$ versus $B$ at three different dephasing strength $\Gamma_d=0.5$, $1$, and $1.5{\rm{eV}}$.
	(e) $R_{37}$ versus the magnetic field $0.03\leq B \leq 0.04$ at three different dephasing strength.
	(f) Two-terminal resistance $R_{15}$ versus the magnetic field $B$
	at two different dephasing strength.
	Each point is only averaged by 50 random configurations because of the small fluctuation shown in Fig. \ref{fig 5}.
	The parameters we use here are given in Appendix \ref{A}.
	}
	\label{fig 4}
\end{figure*}

In addition, the energy distributions in Fig. \ref{fig 3}(b) and (c)
do not show the expected double-step type, even around QPCs.
This is caused by multiple reflections of electrons that have not fully relaxed to equilibrium in the loop of the Fabry-P\'{e}rot interferometer.
In experiments, the edge states that a few micrometers downstream of the QPCs still remain non-equilibrium \cite{le_sueur_energy_2010}, so it is reasonable that the distribution at all positions in Fig. \ref{fig 4}(b) and (c) remains non-equilibrium.
It is also worth mentioning that
the integral of the distribution function in Fig. \ref{fig 3}(b)
is equal at different positions due to the conservation of particle numbers during the propagation without the backscattering,
and so is the integral in Fig. \ref{fig 3}(c).
This indicates that backscattering indeed does not occur, but there exists still dissipation when electrons propagate along the QH edge state.

As a brief summary of Sec. \ref{III}, we point out that
even in the case of topological protection,
non-equilibrium QH edge states will dissipate,
which is the result of inevitable relaxation and entropy increase.
Dissipated heat generation is prohibited only when the electrons
in the edge state are in an equilibrium Fermi distribution.
The coupling between the edge states (e.g., by edge reconstruction \cite{marguerite_imaging_2019} and constrictions \cite{fang_thermal_2021,zhang_dissipation_2020}) will lead to a non-equilibrium, resulting in dissipation at the coupling position and downstream of the coupling position.
Therefore, if one expects to obtain a non-dissipative device,
ensuring that no coupling between chiral edge states flowing from different electrodes with different chemical potentials is needed.

\section{the effect of dephasing in the QH interferometers}\label{IV}

As an interferometer, the most important part is the oscillation curve of
the resistance with the magnetic field (or the gate voltage in experiments).
In this section, we use Eq. (\ref{eq5}) and its boundary conditions to simulate the periodic oscillation of the resistances due to the AB effect at different dephasing strengths.
The results show a competition between the AB interference
and the QH plateau, as well as a proper dephasing strength
can promote the appearance of the QH plateau.
The ideal case where $\Gamma_d=0$, whose coherence length is infinite,
is discussed in Appendix \ref{B},
in which the periodic oscillation of resistance occurs well and the {quantized} plateaus of the Hall resistance $R_{37}$ disappear completely.

\subsection{The interference of resistances}

As mentioned above, QH edge states will be reflected and transmitted by non-integer probability at QPCs, which is similar to the beamsplitters in optics.
With good coherence, after reflection{s}, the electrons accumulate a phase $\Phi=\Phi_{B}+\Phi_d$ each time they go around the interference loop, where $\Phi_{B}$ and $\Phi_d$ are the magnetic phase and dynamical phase, respectively.
In general, the magnetic phase is related to the charge carried by the quasiparticles, which leads to the application of QH interferometers
in the measurement of fractional charges \cite{carrega_anyons_2021}.
However, we do not add the interaction directly to the calculation,
so the magnetic phase gives a periodic $\Delta B=2\pi\phi_0/S$,
where $S$ is the area of the interference loop \cite{halperin_theory_2011,fang_dephasing_2023}.
A more detailed discussion about periods is in Appendix \ref{B}.

As shown in Fig. \ref{fig 4}, both the two-terminal resistance $R_{15}$ and the Hall resistance $R_{37}$ do oscillate with the magnetic field $B$ as expected.
While Hall resistance $R_{28}$ ($R_{46}$) does not oscillate with $B$, since there is no closed interference loop connect lead-2 (lead-4) and lead-8 (lead-6).
When the dephasing strength is small ($\Gamma_d=0.05$), a violent oscillating $R_{37}$ is observed as shown in Fig. \ref{fig 4}(a) and (e).
In this case, the coherence length $L_{\phi}$ is longer than the {perimeter of the interference loop $L_{\rm{loop}}$} and the interference visibility is high, resulting in a high-performance QH interferometer.
As the dephasing strength increases, as shown in Fig. \ref{fig 4}(a-c),
the amplitude of the $R_{37}$ oscillation is obviously suppressed.
In Fig. \ref{fig 4}(e), it can be seen more clearly that
the interference visibility decays with the increase
of the dephasing strength,
while the period of the oscillation remains constant.
This is due to the breakdown of the interference loop
when the coherence length $L_{\phi}$ of the electrons is no longer much larger {$L_{\rm{loop}}$}, which is desired to avoid in experimental QH interferometers.
For example, the presence of a single atomic defect on the physical edge of graphene can cause inelastic scattering leading to dephasing,
which is similar to the virtual leads we introduced.
So moving the edge states away from the physical edge of graphene may improve the interference visibility of the QH interferometer, and some experiments have noted this \cite{ronen_aharonovbohm_2021}.

Naively, as a quantum effect, the QH plateau of the Hall resistance
should be destroyed by dephasing.
However, as shown in Fig. \ref{fig 4}, QH plateaus appear gradually
with the enhancement of dephasing.
Under a small dephasing strength $\Gamma_d=0.05$,
the Hall resistance $R_{37}$ is not quantized.
But under a strong dephasing strength $\Gamma_d=0.5$,
$R_{37}$ reaches the resistance quantum $h/2e^2$ precisely
and the resistance plateaus appear.
This is the result of a competition between the AB interference and the quantized Hall plateau, which can be understood with the relationship between {the perimeter of the interference loop $L_{\rm{loop}}$}, the coherence length $L_{\phi}$, and the magnetic length $L_B$.
When $L_B\ll {L_{\rm{loop}}} \ll L_{\phi}$, the whole interference loop maintains good coherence,
i.e., electrons can be scattered coherently between lead-3 and lead-7 along the interference loop.
As a result, the topological protection is broken and the potential QH plateaus of $R_{37}$ are destroyed by AB oscillations.
With the enhancement of dephasing, $L_{\phi}$ decreases gradually.
When $L_B\ll L_{\phi} \ll {L_{\rm{loop}}}$, the interference loop can no longer realize a visible AB oscillation,
so the amplitude of the oscillation of $R_{37}$ is suppressed. However, quantized Hall plateaus appear.
Here, the Hall resistance is defined as $R_{28}$, $R_{37}$, or $R_{46}$.
Resistances across one or two QPCs, such as $R_{38}$, are not quantized.
Also, if one measures the longitudinal resistance and two electrodes across (not across) a QPC are used to measure the longitudinal voltage, the result will be non-zero (zero).
This is due to the unique geometry of the interferometer, which is different from the conventional Hall bar.
In this parameter regime, the phase memory of electrons in edge states will be rapidly lost during their propagation along the interference loop, resulting in the absence of coherent scattering between lead-3 and lead-7.
At the same time, the strong magnetic field makes the chiral edge state not affected by the local perturbations.
Therefore quantized resistance plateaus of $R_{37}$ appear like the topological protected QH effect.

It is worth noting that with a sufficiently strong dephasing,
$L_{\phi}\lesssim L_B\ll {L_{\rm{loop}}}$, the QH effect can be destroyed again.
As a result, the transition regions between QH plateaus are broadened and QH plateaus are narrowed.
As shown in Fig. \ref{fig 4}(d), the second QH plateau $R_{37}=h/4e^2$ is almost completely destroyed at $\Gamma_d=1.5$.
Meanwhile, a characteristic of classical Hall resistance that varies linearly with the magnetic field appears at the low magnetic field as shown in Fig. \ref{fig 4}(d).
This shows the transition from a quantum insulator to a classical metal caused by strong dephasing, which is consistent with previous theories \cite{pruisken_universal_1988}.

Besides, the interference signal can also be read through the two-terminal resistance $R_{15}$, and it also exhibits periodic oscillations when the filling factor $|\nu|<1$, shown in Fig. \ref{fig 4}(f).
The oscillations of $R_{15}$ and $R_{37}$ are synchronized, which is clearly shown in Fig. \ref{figa}(e) and can be understood by a ballistic transport picture \cite{fang_dephasing_2023}.
With the increase of dephasing strength, the interference visibility of $R_{15}$ decreases for the same reason with $R_{37}$.
However, due to the presence of backscattering at QPCs,
$R_{15}$ will not return to a quantized value after dephasing.

\subsection{The fluctuation of resistances}

Since there are random 1/4 sites coupled to virtual leads in the numerical calculation, it is worth studying how much the random configurations of virtual leads affect the results of resistance.

\begin{figure}
	\centering
	\includegraphics[width=\columnwidth]{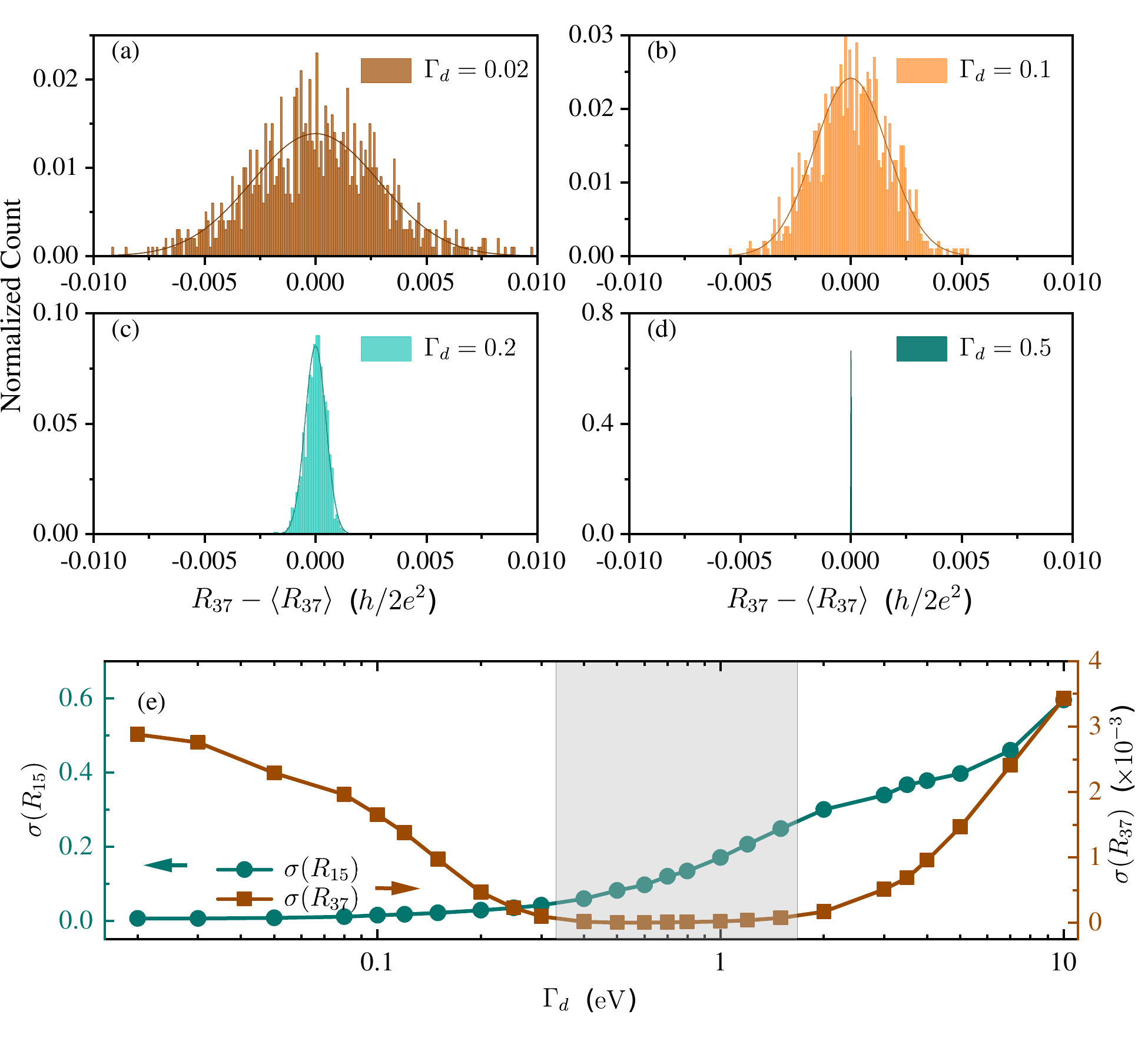}
	\caption{(a-d) The histogram of frequency distribution of $R_{37}-\langle R_{37} \rangle$ at four different dephasing strength $\Gamma_d=0.02, 0.1, 0.2$ and $0.5{\rm{eV}}$ with 1000 random configurations.
	(e) The standard deviation $\sigma(R)=\sqrt{\langle R^2\rangle-{\langle R \rangle}^2}$ of $R_{15}$ and $R_{37}$ vs dephasing strength $\Gamma_d$. The unit of the standard deviation is set to be $h/2e^2$.
	Each point is calculated by 1000 random configurations.
	The shadow marks the area where the Hall resistance $R_{37}$ has a fluctuation close to 0.
	$B=0.03$ and other parameters we use here are given in Appendix \ref{A}.
	}
	\label{fig 5}
\end{figure}

As shown in Fig. \ref{fig 5}, we analyze the fluctuation of resistances
by calculating the deviation from the mean value and the standard deviation of the resistances.
For the Hall resistance $R_{37}$, the normalized count in the histogram shows the form of a Gaussian distribution, and there is a resistance fluctuation $\sigma(R_{37})\approx 3\times10^{-3}(h/2e^2)$ when the dephasing strength is weak $\Gamma_d=0.02$, as shown in Fig. \ref{fig 5}(a).
When the strength of dephasing increases, fluctuations caused by random configurations are significantly suppressed, as shown in Fig. \ref{fig 5} (b-d).
The Hall resistance then enters a plateau of zero fluctuations,
with standard deviation $\sigma(R_{37})=0$ [around $10^{-5}\sim10^{-6}(h/2e^2)$], as shown in the shadow region of Fig. \ref{fig 5}(e).
This indicates that the dephasing can promote the appearance of the QH effect.

These results provide another way to understand
the competition between AB interference and topologically protected QH resistance.
When the dephasing strength is weak, the coherence length is much larger than the size of the QH interferometer with $L_B\ll L_{\rm{loop}} \ll L_{\phi}$,
AB interference dominates the transport process,
as shown in Fig. \ref{fig 4}.
The top and bottom QH edge states can be connected by backscattering at two QPCs forming a coherent quantum state along the interference loop, which causes the breakdown of the topological protection.
Therefore, there is a strong resistance fluctuation.
On the contrary, when the dephasing effect is strong,
the coherence length $L_{\phi}$ is much smaller than the perimeter of the interference loop with $L_B\ll L_{\phi} \ll L_{\rm{loop}}$.
Coherent scatterings of electrons between lead-3 and lead-7 disappear.
On the one hand, this is accompanied by the appearance of the quantized Hall plateau, as we discussed above [see Fig. \ref{fig 4}(e)].
On the other hand, the robustness of topological protection to perturbations is also shown, and the Hall resistance fluctuation $\sigma(R_{37})$ is limited to 0, as shown in Fig. \ref{fig 5}(d) and (e).
Therefore, we also see that the QH effect is promoted by proper dephasing from the fluctuation of resistance.

As the dephasing effect continues to increase
and the coherence length $L_{\phi}$ decreases to nearly
the same order of the magnetic length (i.e. $L_{\phi}\lesssim L_B\ll L_{\rm{loop}}$),
the topological protection is broken by dephasing and
the QH effect no longer occurs, as we discussed above.
This causes the fluctuation of the Hall resistance $R_{37}$
at the right of the shaded region in Fig. \ref{fig 5}(e) to increase again.
For the two-terminal resistance $R_{15}$, without the topological protection, its fluctuation increases monotonically
with the increase of the dephasing strength.

\section{Conclusion}\label{V}

It is usually assumed that QH edge states do not dissipate,
but recent theoretical and experimental works have challenged this stereotype.
According to numerical calculations in the QH interferometer,
inevitable dissipation will occur in the QH edge states,
accompanied by heat and entropy generation.
By calculating the energy distribution of electrons, we find that this dissipation comes from the relaxation of the edge state from non-equilibrium to equilibrium.
Only equilibrium, not topological protection, can prevent dissipation.

Furthermore, our numerical results also simulate
the dephasing process of QH interferometers from complete coherence.
The amplitude of resistance oscillation caused by AB interference decreases with the increase of dephasing strength and
proper dephasing strength promotes the appearance of the QH plateaus.
The competitive process of the quantum oscillation and topologically protected QH plateaus is shown in terms of resistances and
the fluctuation of resistances.
We explained it clearly with the relationship
between the coherence length $L_{\phi}$ and the perimeter $L_{\rm{loop}}$ of the interference loop.

In general, our work contributes in two ways.
(1) In the design of the non-dissipative device,
we need to keep the QH edge states of the device at equilibrium
all along while working.
(2) We point out that the QH interferometer is a promising platform for
studying the relaxation processes from non-equilibrium to equilibrium
and the relationship between dissipation and the topological protection.

\section*{ACKNOWLEDGMENTS}
P.-Y. Liu is grateful to J.-Y. Fang for fruitful discussions.
This work was financially supported by the National Natural Science Foundation of China (Grant No. 12374034 and No. 11921005), the Innovation Program for Quantum Science and Technology
(2021ZD0302403), and the Strategic priority Research
Program of Chinese Academy of Sciences (Grant No.
XDB28000000). We also acknowledge the Highperformance
Computing Platform of Peking University for
providing computational resources.\\

\appendix

\section{The parameters we use}\label{A}

In this appendix, we give all the parameters used in the calculations.

\begin{figure}[!htb]
	\centerline{\includegraphics[width=\columnwidth]{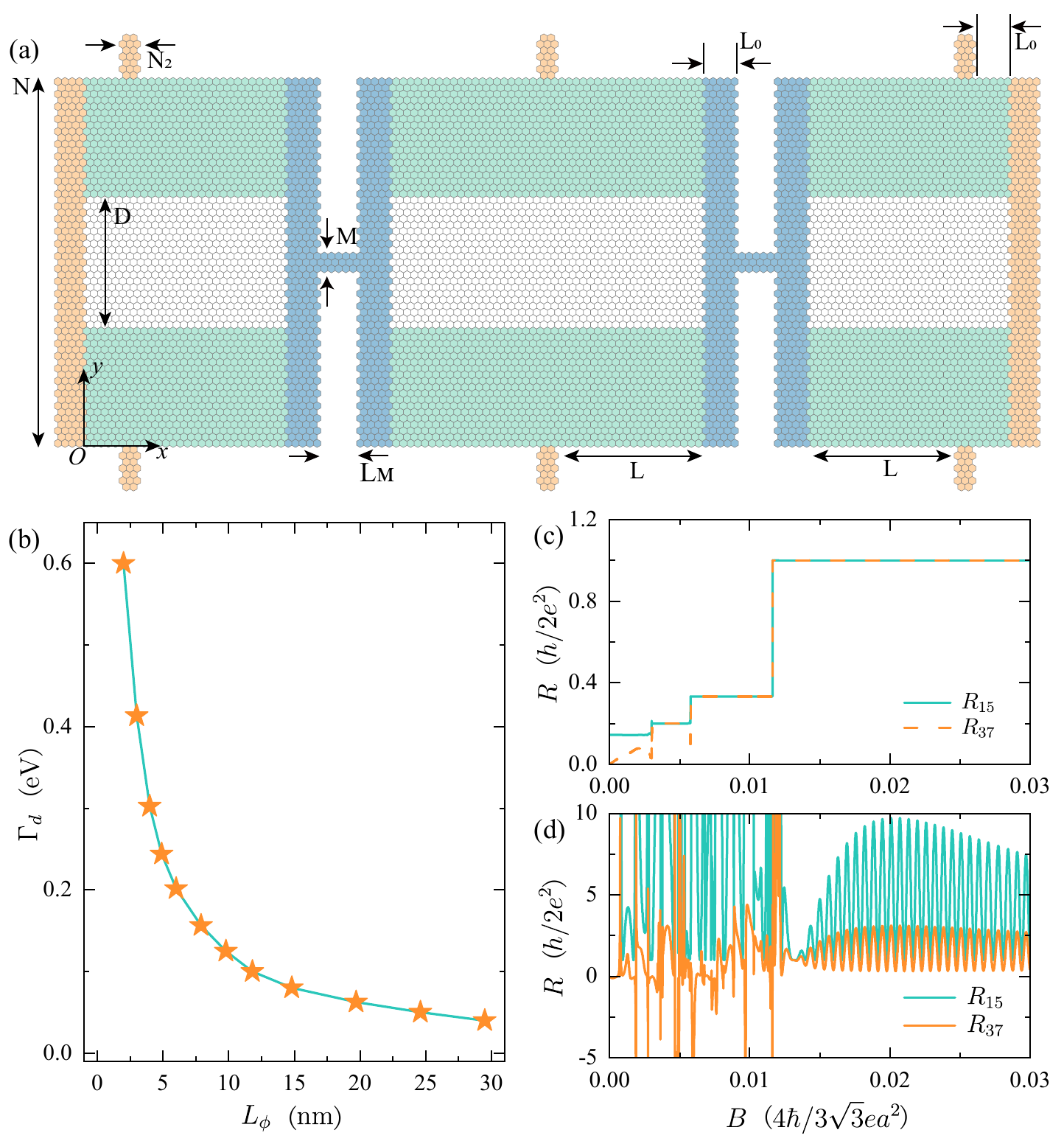}}
	\caption{(a) A schematic diagram that gives detailed parameters of the QH interferometer. The dimensions parameter and the coordinates of the system are indicated on the diagram.
	(b) The the conversion relationship between $L_{\phi}$ and $\Gamma_d$.
	(c) The two-terminal resistance $R_{15}$ and the Hall resistance $R_{37}$ versus the magnetic field $B$, without the presence of QPCs and dephasing.
	(d) $R_{15}$ and $R_{37}$ versus $B$, with the presence of QPCs and the absence of dephasing.
	The parameters we used in (b-d) are given in Appendix \ref{A}.}
	\label{figa}
\end{figure}

First, the geometric details of the QH interferometer are shown in Fig. \ref{figa}(a), where $N=89$, $N_2=3\sqrt{3}$, $M=5$, $L=20\sqrt{3}$, and $L_0=L_M=5\sqrt{3}$ in the unit of the length of carbon-carbon bond $a=0.142{\rm{nm}}$.
In the problem we discussed, the transport is carried by the QH edge state while the bulk is insulated.
So we only couple virtual leads in the green and blue regions of the device, leaving an area of width $D=31$.
In real calculations, we use $L=40\sqrt{3}$, $L_0=L_M=10\sqrt{3}$ and other parameters are same with those in Fig. \ref{figa}(a).
The perimeter of the interference loop $L_{\rm{loop}}\approx 2N+4L+4L_0+2N_2$.
In the calculation of the whole paper, unless specifically mentioned, we do not change these size parameters.
Besides, a coordinate is established to describe all sites \cite{fang_thermal_2021}. The origin point is located in the bottom left corner and the QH interferometer is in the $xOy$ plane,
as shown in Fig. \ref{figa}(a).
	Virtual leads are coupled to the QH interferometer randomly.
When we calculate the heat generation (Fig. \ref{fig 2}), the interference (Fig. \ref{fig 4}) and the fluctuation (Fig. \ref{fig 5}) of resistances, we randomly couple virtual leads to some sites in the green and blue regions in Fig. \ref{figa}(a).
However, when we calculate the energy distribution of electrons (Fig. \ref{fig 3}), virtual leads are coupled only in the green regions.
Each site in the selected regions has a probability $\eta=1/4$ to couple a virtual lead.
Our results are averaged under a number of random configurations of virtual leads.

Second, considering a magnetic field $B$ perpendicular to the paper
facing outward, we give the gauge conventions used in calculations.
In principle, choosing different gauges does not affect the calculation result.
Whatever gauge is chosen, each hexagonal unit should have a magnetic flux $2\phi=3\sqrt{3}a^2B/2\phi_0$, where $\phi_0=\hbar/e$ is the flux quantum, and the unit of the magnetic field is given by $4\hbar/3\sqrt{3}ea^2$.
But to obey the translation symmetry of leads, we choose $\bm{A}=(-By,0,0)$ in the QH interferometer, lead-1, and lead-5.
For lead-(2-4) and lead-(6-8) we choose $\bm{A}=(0,Bx,0)$ \cite{fang_thermal_2021}.
With this gauge convention, the Peierls phase $\phi_{ij}$ in Eq. (\ref{eq1}) can be calculated \cite{long_disorder_2008,fang_thermal_2021}.

At last, we convert the dephasing strength $\Gamma_d$
into the coherence length $L_{\phi}$, which is an observable \cite{roulleau_direct_2008}.
We use a zigzag graphene nanoribbon perfectly contacted with two real left and right leads (labeled by $L$ and $R$) \cite{xing_influence_2008,fang_dephasing_2023}.
This nanoribbon has a y-direction width $N=89$, $B=0.03$ (the same parameters as our QH interferometer), and is also coupled by virtual leads (labeled by $\nu=1,2,3,\dots$) with $\eta=1/4$.
The transmission coefficients from lead-$p$ to lead-$q$ $T_{pq}(E=0)$ can be calculated with the same method we described in Sec. \ref{II}.
Here, $p$ and $q$ can be both real leads and virtual leads $(p,q=L,R,1,2,3,\dots)$.
For example, $T_{L\nu}$ is the transmission coefficient from the real left lead to the $\nu$-th virtual lead.
It is clear that all these transmission coefficients are functions of the nanoribbon length in x-direction $L_x$ and $\Gamma_d$ [$T_{pq}=T_{pq}(L_x,\Gamma_d)$].
$T_{LR}$ represents the probability that electrons go through the nanoribbon without dephasing by virtual leads.
The summation of transmission coefficients from the real left lead to all the virtual leads $\sum_{v=1,2,3,\dots}T_{Lv}$ represents the probability that electrons lose their phase information.
If we fix $L_x$ and increase $\Gamma_d$, $T_{LR}$ will decrease and $\sum_{v}T_{Lv}$ will increase.
When $T_{LR}(L_x,\Gamma_d)=\sum_{v}T_{Lv}(L_x,\Gamma_d)$, which means the probability of an electron from the left lead losing and keeping phase information are both $50\%$, $L_x=L_{\phi}$.
Therefore, the conversion relationship between $L_{\phi}$ and $\Gamma_d$ is given by $T_{LR}(L_{\phi},\Gamma_d)=\sum_{v}T_{Lv}(L_{\phi},\Gamma_d)$, which can be numerically calculated as shown in Fig. \ref{figa}(b).

\section{The case with infinite coherence length}\label{B}

With Eq. (\ref{eq5}) and the boundary conditions we set in Sec. \ref{II}, we can calculate resistances without the presence of virtual leads $\Gamma_d=0$.
In other words, we can consider the ideal case where the coherence length of the system is infinite.

First, we calculate the resistances without the presence of QPCs, i.e., $M=N=89$.
As shown in Fig. \ref{figa} (c), the two-terminal resistance $R_{15}$ and the Hall resistance $R_{37}$ are quantized precisely as $\frac{h}{2e^2}\times (1,\frac{1}{3},\frac{1}{5})$, which is consistent with the QH effect of graphene \cite{novoselov_two-dimensional_2005}.
The number of Hall plateaus depends on the on-site energy $\epsilon_i$ and the width $N$.

Further, with two QPCs, the Hall bar becomes a Fabry-P\'{e}rot QH interferometer finally.
The absence of dephasing results in an infinitely long coherence length.
As a result, both $R_{15}$ and $R_{37}$ oscillate violently, and the QH plateaus disappear, as shown in Fig. \ref{figa} (d).
When $B < 0.0116$, the interference between different edge states leads to a random oscillation \cite{fang_dephasing_2023}.
When $B > 0.0116$, there is only one edge state, showing periodic oscillations of $R_{15}$ and $R_{37}$ in Fig. \ref{figa}(d),
which is nothing but the period of the Fabry-P\'{e}rot interference \cite{carrega_anyons_2021}.
As we mentioned in Sec. \ref{IV}, the period of the oscillation is given by $\Delta B=2\pi \phi_0/S$.
If we estimate $S$ as the total area of cells between two QPCs, the theoretical period is $\Delta B=\frac{2\pi \phi_0}{(3\sqrt{3}a^2/2)\times(30\times103+29\times102)}=5.19\times10^{-4}$, in the unit of $4\hbar/3\sqrt{3}ea^2$.
We can also use FFT to extract the period from the oscillations, $\Delta B=6.09\times10^{-4}$, which is slightly larger than the theoretical one.
This is because the area of the interference loop is smaller
than the area between the two QPCs.

\bibliography{ms.bib}

\end{document}


\renewcommand\thesection{S\Roman{section}}
\def\theequation{S\arabic{equation}}
\def\thefigure{S\arabic{figure}}

\title{Supplementary materials for ``Dissipation and dephasing in quantum Hall interferometers"}

\author{Peng-Yi Liu}
\affiliation{International Center for Quantum Materials, School of Physics, Peking University, Beijing 100871, China}
\author{Qing-Feng Sun}
\email[]{sunqf@pku.edu.cn}
\affiliation{International Center for Quantum Materials, School of Physics, Peking University, Beijing 100871, China}
\affiliation{Hefei National Laboratory, Hefei 230088, China}
\date{\today}

\maketitle


In the main text, we use B\"{u}ttiker virtual leads to simulate the dissipation and dephasing effect.
The coupling between virtual leads and the quantum Hall interferometer is described by the last term of Eq. (2), which is given by $\sum_{i,k}(t_{ik}a_{ik}^\dagger c_i+H.c.)$.
Here, $t_{ik}$ is randomly either 0 or $t_k$ with the probability $1-\eta$ and $\eta=1/4$.
Specifically, we determine the coupling strength at each site using a uniformly random number between $0$ and $1$.
$t_{ik}$ will change abruptly from $t_k$ to $0$ if the random number transitions from below to above $\eta$.
It is conceivable that allowing $t_{ik}$ to take on a wider range of values, perhaps following a continuous distribution, could yield more realistic simulations.
Consequently, we perform a comparative calculation to validate the robustness and reliability of our approach in the main text.

\begin{figure}[htbp]
	\includegraphics[width=0.7\columnwidth]{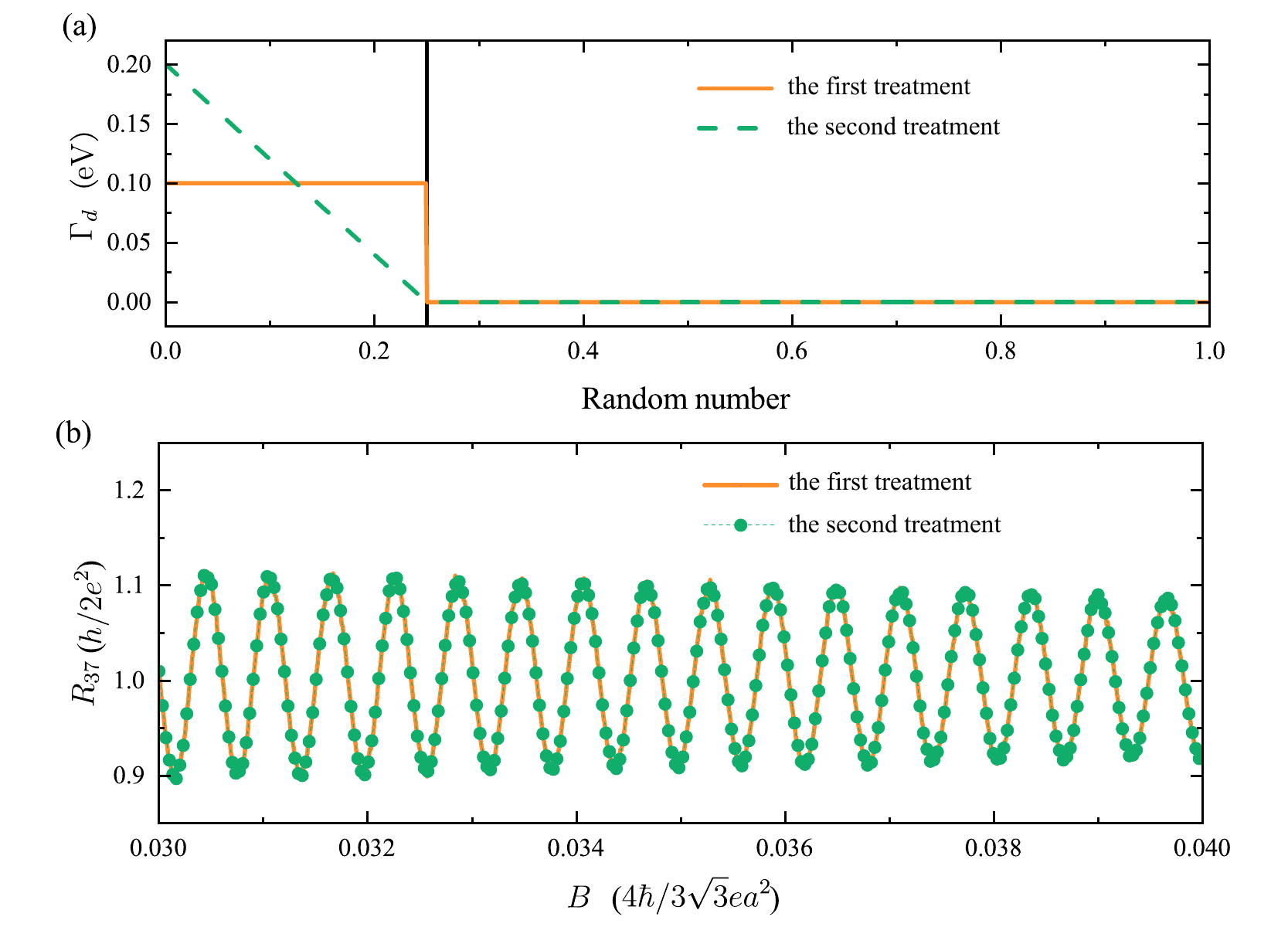}
	\caption{\label{FIGS1}
	\baselineskip 14pt
	\sl	
	A comparative calculation about two different treatments of the coupling of virtual leads.
	(a) The abruptly changed $\Gamma_d$ (orange solid curve) with the random number in the first treatment and the continuously changed $\Gamma_d$ (green dashed curve) in the second treatment. 
	The black solid line shows $\eta=1/4$.
	(b) The result of the first treatment, i.e. the orange curve in Fig. 4(e) is shown by the orange solid curve.
	The result of the second treatment is shown by green dots.
	All the parameters are the same as in Fig. 4(e) of the main text except $\Gamma_d$.
}
\end{figure}

We employ two distinct treatments for $t_{ik}$ in calculating $R_{37}$, while keeping all other parameters unchanged.
The first treatment is our approach in the main text, and the result is the orange curve ($\Gamma_d=0.1$) in Fig. 4(e) of the main text.
In the second treatment, we assign a probability of $1-\eta$ for $t_{ik}^2$ to be exactly 0, and a probability of $\eta$ for $t_{ik}^2$ to be uniformly distributed within the interval $[0,2t_k^2]$. 
This setup allows the linewidth function of the virtual leads $\Gamma_d$, to become a continuous random variable rather than a discrete one, as depicted in Fig. \ref{FIGS1}(a).
After averaging over 50 random configurations of the virtual leads, we find that both treatments yield virtually identical results [Fig. \ref{FIGS1}(b)], with a difference of less than 0.01, indicating that our treatment used in the main text is reliable and that the results are insensitive to the specific details of the coupling.

